\documentclass{article}
\usepackage{arxiv}

\usepackage[utf8]{inputenc} % allow utf-8 input
\usepackage[T1]{fontenc}    % use 8-bit T1 fonts
\usepackage{hyperref}       % hyperlinks
\usepackage{url}            % simple URL typesetting
\usepackage{booktabs}       % professional-quality tables
\usepackage{amsfonts}       % blackboard math symbols
\usepackage{nicefrac}       % compact symbols for 1/2, etc.
\usepackage{microtype}      % microtypography
\usepackage{lipsum}		% Can be removed after putting your text content
\usepackage{graphicx}
\usepackage{natbib}
\usepackage{doi}

\begin{document}

\title{Everywhere \& Nowhere: Envisioning a Computing Continuum for Science}

\author{\href{https://orcid.org/0000-0003-0983-7408}{\includegraphics[scale=0.06]{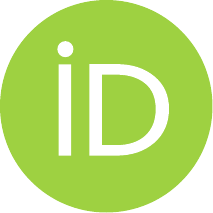}\hspace{1mm}Manish Parashar}\thanks{{\em Manish Parashar} is Director of the Scientific Computing and Imaging (SCI) Institute, Chair in Computational Science and Engineering, and Presidential Professor, Kalhert School of Computing at the University of Utah. Manish is the founding chair of the IEEE Technical Community on High Performance Computing (TCHPC), and is Fellow of AAAS, ACM, and IEEE/IEEE Computer Society. For more information, please visit \url{http://manishparashar.org}.}\\
	Scientific Computing and Imaging (SCI) Institute, \\
 University of Utah, \\
 Salt Lake City, UT, 84112, USA \\
	\texttt{manish.parashar@utah.edu} \\
}

\maketitle

\begin{abstract}
Emerging data-driven scientific workflows are seeking to leverage distributed data sources to understand end-to-end phenomena, drive experimentation, and facilitate important decision-making. Despite the exponential growth of available digital data sources at the edge, and the ubiquity of non trivial computational power for processing this data, realizing such science workflows remains challenging. This paper  explores a computing continuum that is everywhere and nowhere -- one spanning resources at the edges, in the core and in between, and providing abstractions that can be harnessed to support science. It also introduces recent research in programming abstractions that can express what data should be processed and when and where it should be processed, and autonomic middleware services that automate the discovery of resources and the orchestration of computations across these resources.~\footnote{This paper is based on the author's IEEE Sidney Fernbach award presentation at SC23, The International Conference for High Performance Computing, Networking Storage and Analysis, Denver, CO, USA, November 2023. (\url{https://sc23.supercomputing.org/})} 
\end{abstract}

\section{Everywhere \& Nowhere: The Emerging Computing Continuum}
Disruptive innovations coupled with  technological advances across the computing stack are resulting in dramatic changes in all aspects of the scientific computing/high-performance computing landscape (HPC). These changes include the emergence of novel processor and system architectures influenced by the end of Dennard scaling and the slowing of Moore's law, and driven by in large part by the transcendence of AI and the dominance of the hyperscalers~\cite{reed2022reinventing}; pervasive availability of non trivial (and growing) computing capacities at the edge, within the networks and along the data-path, and extreme capabilities at high-performance computing and cloud data centers; rapidly growing data sources and data volumes and rates; and increasing network bandwidths and in-network services. 
%It is also interesting how these innovations are largely delivered, as services rather than as products. 

 These changes are coupled with a shift in the workloads that are driving the innovations. These workloads are increasingly data-driven, leverage artificial intelligence/machine learning (AI/ML) techniques, and have changing and flexible notions of precision (i.e., exploring 8 and 16 bit, in addition to the typical 64 bit). The supporting software stacks for resource access and application development, deployment, and execution are similarly evolving to include, for example, as-a-service, serverless, and containerized approaches that are typically found in enterprise systems. 
 
 Perhaps the most significant change is in the value structures, i.e., what users value most. Maximizing performance was traditionally valued above all else, but we are increasingly seeing that other aspects, such as ease of access and use, time to science, energy/environmental impact, are becoming more important to users, and users are willing to give up some performance in order to gain in these dimensions. 

Clay Christensen, in his book \emph{Innovators Dilemma~\cite{innovators-dilemma}}, explored the impacts of such changes in value structures on innovation and technology. He noted that, when what the customers value changes, it creates opportunities for disruptive innovations allowing for the introduction of new technologies and approaches. We may be experiencing this in HPC, where existing performance levels, on average, meet user needs, and the differentiators are along other attributes such as usability, accessibility, robustness, environmental impacts, etc. At the same time, there are new applications, such as AI/ML-based workflows, that are driving alternate solutions across the stack. Looking at the current HPC landscape, it is clear that what applications are driving it, and how these applications use it, are evolving very quickly.

The result of these technological (and socio-technical) innovations and disruptions is a rapidly emerging, connected, and seamlessly accessible continuum of computational (computing, data, communication) capabilities~\cite{balouek2019towards}. Furthermore, a vision of HPC enabled by this continuum is one that is {\emph{everywhere and nowhere}.

We are already seeing the pervasive integration and availability of non trivial computing capabilities everywhere, in our automobiles, in devices that we carry, in appliances that we use, etc. HPC is integrating itself in some form or another into all aspects of our lives, and these HPC capabilities will only increase as technology advances -- i.e., \emph{HPC is everywhere}. 

 At the same time, how we interact with and use HPC is also changing. Traditionally, using HPC has been an involved process -- we develop specialized codes, which are compiled on a special system, and then queued to run on the HPC system -- i.e., a very deliberate process often involving multiple systems. This process is in contrast with current trends across enterprise software stacks and the abstractions they provide to the user, e.g., those based on services, containers, serverless deployments, and notebook/gateway access. As the HPC community increasingly embraces these abstractions, we are seeing a move to a more transparent way of using HPC that is more seamlessly integrated into our workflows and our lives: i.e., \emph{HPC will be everywhere and nowhere.}

\section{Harnessing the Computing Continuum for Science}

\begin{figure}
\centerline{\includegraphics[width=30pc]{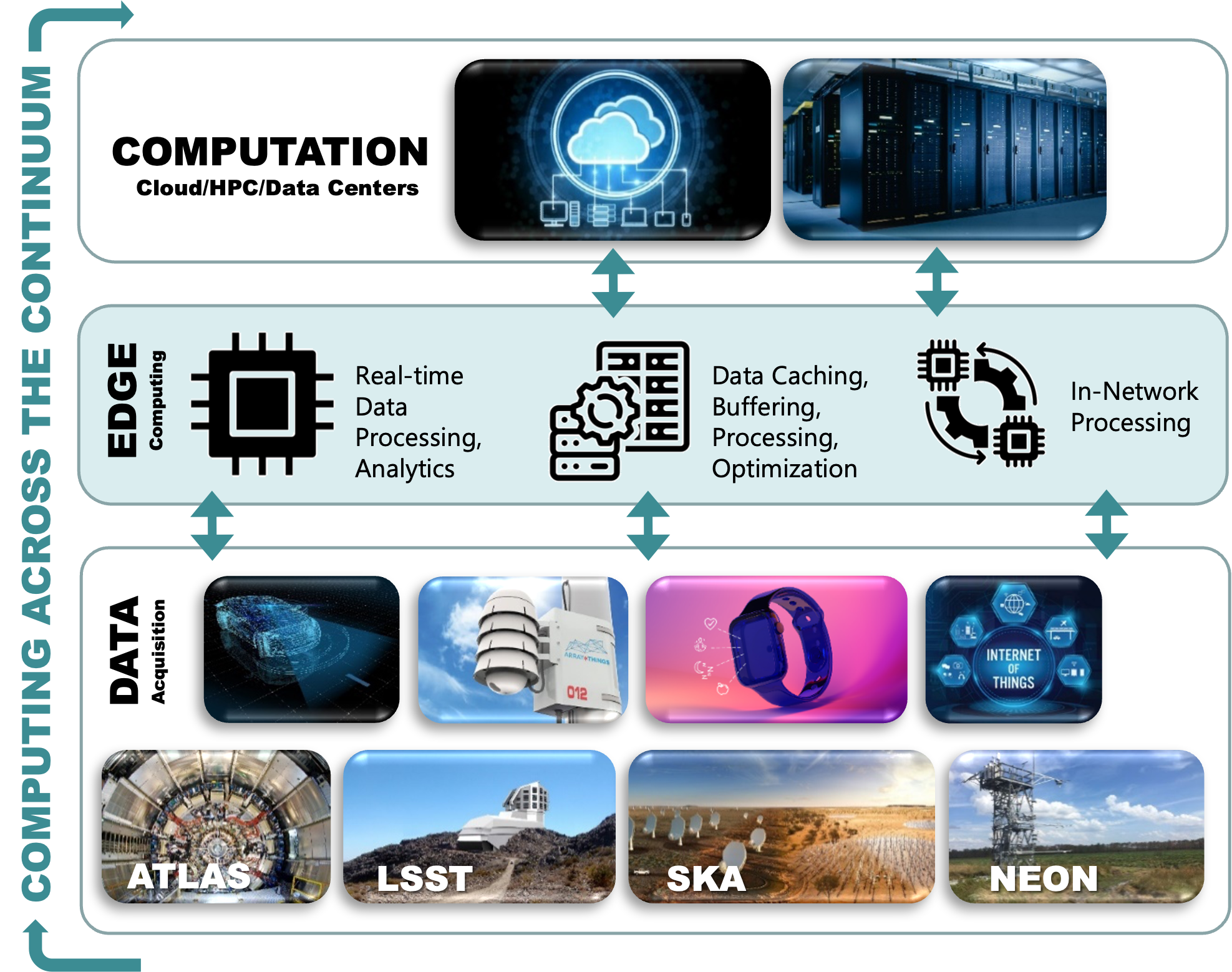}}
\caption{A computing continuum across the evolving science ecosystem spanning large-scale instruments, experimental facilities, observatories, and sensor networks, all streaming data; high-speed networks and advanced network services; and a range of computing capabilities and capacities along the continuum, from edge, to in-network, to large-scale data centers.}\vspace*{-5pt}
\label{fig:science-continuum}
\end{figure}

There exists a natural continuum across the evolving science ecosystem spanning large-scale instruments, experimental facilities, observatories, and sensor networks, all streaming data; high-speed networks and network services; and a range of computing capabilities along the continuum, from edge, to in-network, to large-scale data centers (see Figure~\ref{fig:science-continuum}). 

This continuum is also spurring a natural evolution in the types of application workflows that are being developed. These workflows combine sensing and streaming data (e.g., from observatories and experimental facilities) with simulations and data-driven modeling and actuation, to understand, analyze, predict and actuate. 

One class of such applications workflows that is enabled by the continuum and is being increasing deployed is end-to-end experiment management, where streaming data from an experiment or instrument is analyzed and modeled, and the result of the modeling is used to control, manage, and/or optimize the experiment. One example here is an instrumented oil-field workflow. The goal of this workflow is to use streaming data from an actual oilfield along with subsurface flow simulations and ML-based optimizers, to manage oil production, reduce environment impacts, etc.~\cite{bangerth2005autonomic}. Another example is from a fusion workflow, and implements a diagnostics system within the continuum that enables early prediction of anomalies while the tokamak experimental facility operates, which is important as these anomalies can damage the instrument\footnote{\url{WDMAPP – The First Simulation Software in Fusion History to Couple Tokamak Core to Edge Physics, https://www.exascaleproject.org/highlight/wdmapp-the-first-simulation-software-in-fusion-history-to-couple-tokamak-core-to-edge-physics/}}.

Another class of applications that can be viewed as a natural extension, is digital twins for large-scale complex system. These systems are digital representations of actual real-world physical systems and can serve as vehicles for understanding, managing, optimizing, protecting, etc., the physical systems~\footnote{National Academies of Sciences, Engineering, and Medicine -- Foundational Research Gaps and Future Directions for Digital Twins, \url{https://nap.nationalacademies.org/catalog/26894/foundational-research-gaps-and-future-directions-for-digital-twins}.}. These applications  highlight the need for combining real-time data acquisition with large-scale modelling, both data-driven and mathematical, and possible actuation, and the computing continuum will play a large role in making these systems a reality.

\section{Urgent Computing and the Computing Continuum}
An important class of applications enabled by the computing continuum is urgent computing. Urgent computing can be defined as computing under strict time and quality constraints to support decision-making with the desired confidence, and within a defined time interval~\cite{balouek2020harnessing}. The goal is to leverage data and computations to support decision-making during an emergency. Figure~\ref{fig:urgent-workflow} illustrated the urgent computing workflow, which uses the computing continuum to process data from a range of data sources along with other resources and services along the continuum to detect events, develop a response, and trigger actions.

\begin{figure}
\centerline{\includegraphics[width=30pc]{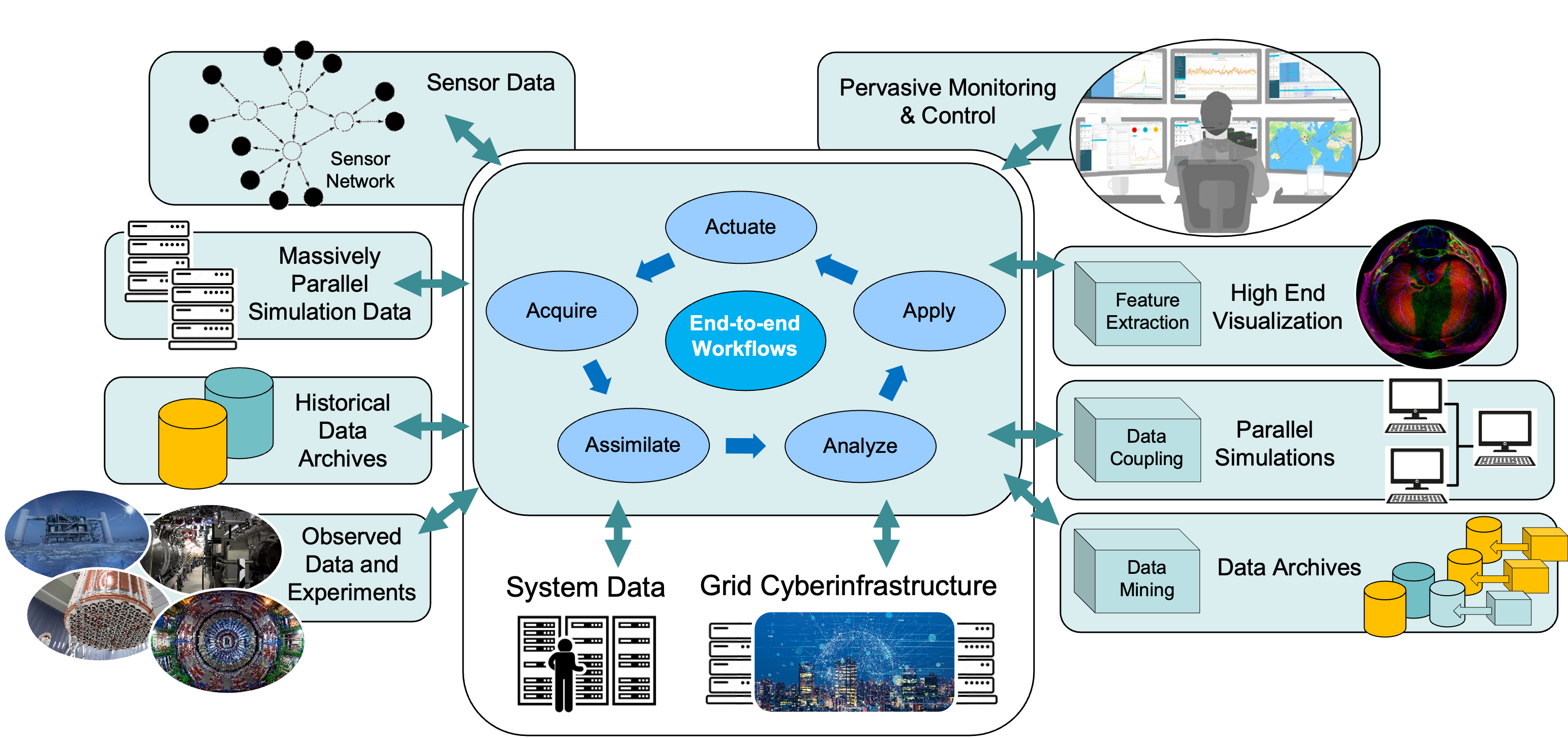}}
\vspace*{5pt}
\centerline{\includegraphics[width=30pc]{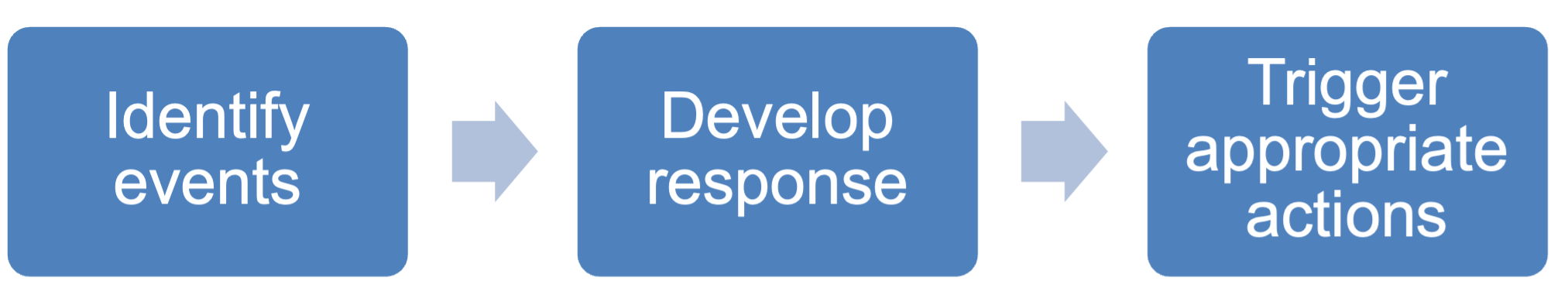}}
\caption{The urgent computing workflow uses the computing continuum to process data from a range of data sources, along with other resources and services along the continuum, to detect events, develop a response, and trigger actions.}\vspace*{-5pt}
\label{fig:urgent-workflow}
\end{figure}

Recent conversations about the importance and potential of urgent computing were triggered by experiences with the COVID19 HPC Consortium~\cite{brase2022covid} during the recent pandemic. The consortium was a remarkable international partnership across government, academia, and industry aimed at providing computing resources for pandemic-related research. It brought together an international group of over 40 resource providers and supported over 100 research projects, leading to important research outcomes spanning drug design, treatments, vaccine research, logistics, etc. 

Although the consortium was very effective in highlighting the tremendous potential of computing and data in dealing with emergencies such as pandemics, extreme weather events, and wildfires, it also highlighted many infrastructure, partnership, training, and policy gaps that prevent these resources and efforts such as the consortium  from achieving the desired impacts in an urgent situation, and the need for thinking differently about HPC, both technically and operationally. 

A more immediate (and local) use case that is driving our current work in urgent computing is  the impact of wildfires on air quality, and specifically, the impact of wildfires in California on air quality in the Salt Lake Valley in Utah. While the impact of wildfires on air quality is a broader issue, there are  significant implications for the Salt Lake Valley (in Utah, USA) due to periodic temperature inversions, which trap cold air underneath a layer of warm air. Such inversions act like a ``lid'' on the Salt Lake Valley, causing particulate pollution to double quickly. 

Recent studies have noted that air pollution is a leading health risk factor globally and has a wide range of health implications, including on physical, respiratory, and mental health. Understanding and predicting these impacts requires combining a range of real-time data feeds about the fires, smoke, wind/weather conditions, and integrating atmospheric models, fire models, air quality models, etc. Furthermore, any action must include an understanding of  the layout of the region, population distributions and demographics, road and traffic conditions, and other similar aspects. Note that air quality can deteriorate to severely unhealthy levels very quickly, making it an urgent challenge, and its impacts are not uniform, typically impacting under-resourced and under-developed neighborhoods the hardest. 

Another example urgent computing use case driving our research is \emph{Early Earthquake Warning.} In this use case, we used machine learning to analyze streaming 3-D time-series data along the continuum, at the edge and in-transit within the network~\cite{fauvel2020distributed}. The goal here is to deliver alerts before the ground motion reaches sensitive areas. 

Technically, ground motion in an earthquake is caused by two types of waves, Primary waves (or P-waves) and Secondary waves (or S-waves). P-waves move longitudinally along the Earth's crust and are around 1.7x faster than S-waves, which move in the transverse direction along Earth's interior. It is the S-waves that are responsible for severe damage. The goal of our research is to detect P-waves as quickly as possible, and it is important to do this as the data streams in. It is also important to combine 3D time series data from seismometers and GPS sensors to detect a range of earthquakes. Seismometers can effectively detect medium (5 $\leq$ magnitude $<$ 6, Richter scale) earthquakes but are not as effective for large earthquakes (6 $\leq$ magnitude, Richter scale), for which GPS sensors are more effective. In this project, we leveraged the computing continuum to integrate and analyze these data streams at the edge and in transit and combined  with modeling results coming from the core to predict the magnitude of seismic events, thereby allowing for timely alerts.

\subsection{Research Challenges}
There are many underlying research challenges in making urgent computing and the underlying data-driven workflows and the computing continuum a reality. Some key research questions include: (1) How do you drive computation through data? (2) How do you ensure security, privacy and trust in all aspects of the infrastructure? (3) How do you accommodate uncertainties in data and computation? (4) How do you build applications and manage workflows so that they can adapt to increase their value? (5) How do you continuously optimize workflow execution in a dynamic data-driven environment? (6) How do you develop robust system infrastructure and services to support dynamic execution? (7) How do you incorporate appropriate utility models, market models, social/trust models, etc.? and (8) How do you formulate the necessary policy and governance structures to manage operation?

Complementing fundamental advances in addressing these research questions is translational research that is critically important in achieving the urgent computing  vision. Translational computer science research\cite{abramson2019translational}\footnote{Translational Computer Science, \url{https://translational-cs.org/}.} is the bi-directional integration and interplay between foundational research and the delivery and deployment of its outcomes. It aims to  closely couple cycles of innovation
between computer science and other disciplines to significantly accelerate the transformative impact of computer science.

\subsection{Recent Research Efforts}
Our recent research addresses some  of the underlying research issues that were listed above. These include:
 
\paragraph{How do you drive computation through data?} Urgent workflows are triggered by attributes and/or content of data streams, and the data attributes/content determines what, when, and where to execute computations. For example, the online analysis of 3D time series data from seismometers and GPS sensors triggers earthquake detection workflows on edge and/or  cloud/HPC resources. The R-Pulsar programming system~\cite{renart2019edge} leverages the Associative Rendezvous (AR) interaction model to allow users to programmatically define data-driven  workflows executing across the computing continuum as reactive behaviors based on the content of streaming data.  It provides abstractions to express workflow topologies that are triggered based on the availability of resources and/or data as well as data values,  statistical trends over time/spatial windows, etc., i.e., data streams are evaluated at runtime to decide when, how, and where to process their data.

\paragraph{How do you discover and aggregate resources based on current needs?} Discovering, federating, and utilizing (computing, data, etc.) resources along the computing continuum in an online manner are essential to urgent computing, including adapting the federation and discovering and aggregating new resources as the application needs evolve and/or new resources become available. Our research has leveraged constraints-based autonomic federation to realize a dynamic software-define system that can support urgent workflows~\cite{abdelbaky2018software}. We have also developed recommendation systems for scientific data to support intelligent data discovery and delivery~\cite{parashar2023toward}. Specifically, this research 1) uses user query analysis techniques that model access patterns and associated localities and affinities; 2) optimizes data caching, data prefetching, and data steaming mechanisms to support optimized push-based data delivery; and 3) develops a data recommendation framework based on the collaborative knowledge-aware graph-attention network (CKAT) recommendation model.

\paragraph{How do you manage execution (and QoS) in a dynamic environment?} The highly dynamic requirements of urgent application workflows coupled with inherent dynamism of the computing continuum warrant autonomic runtimes that can effectively manage and optimize execution. Our research is developing autonomic runtime services for data and workflow management as part of the Virtual Data Collaboratory and National Data Platform~\cite{parashar2023toward} projects. Specfically, these services manage workflow  scheduling and execution across the computing continuum based on user-defined policies and constraints to implement necessary tradeoffs.

\section{Conclusion: A Call to Action}

\begin{figure}
\centerline{\includegraphics[width=35pc]{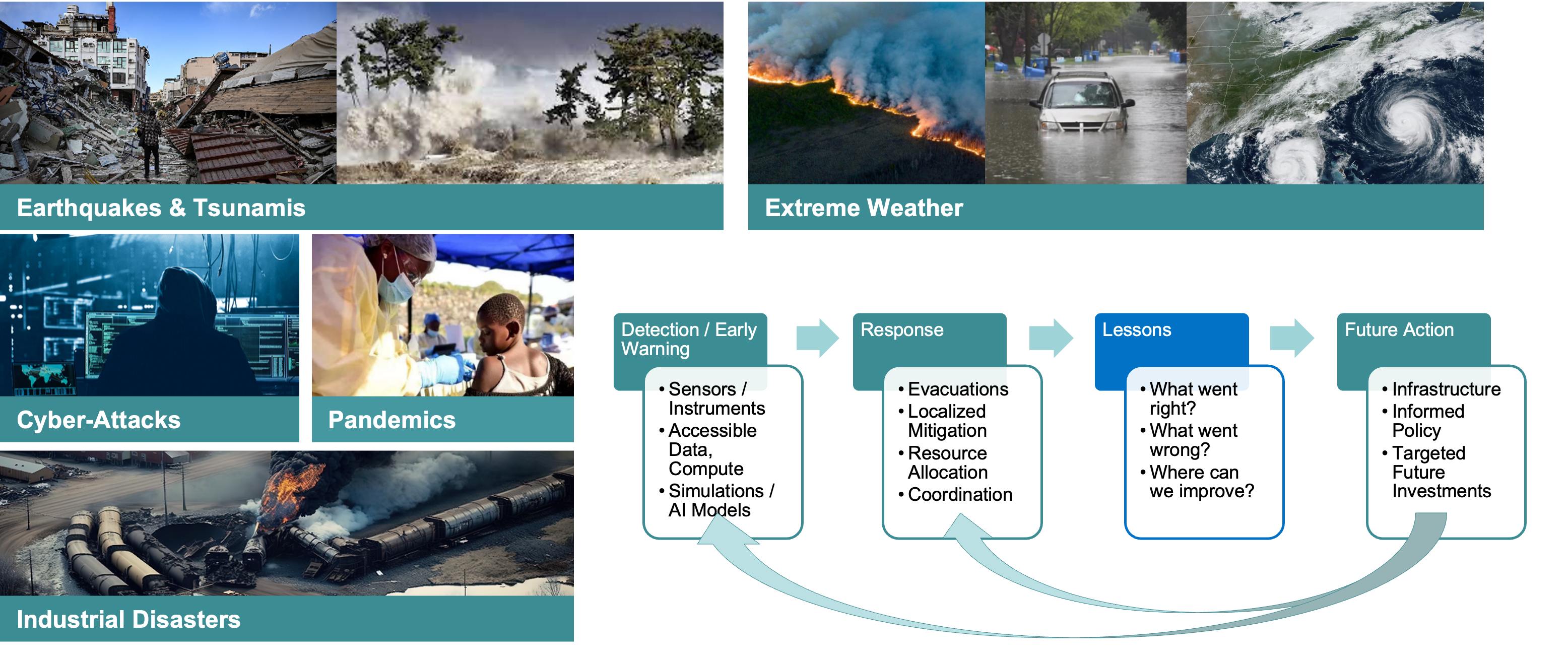}}
\caption{A call to action: We are witnessing urgent events with increasing frequency and increasing impacts, and the computing continuum along with urgent computing has the potential to help us understand, manage, and mitigate the impacts of these event. The HPC community has an opportunity to collectively leverage its expertise and the computing continuum to make a difference.}\vspace*{-5pt}
\label{fig:calltoaction}
\end{figure}

We are witnessing urgent events with increasing frequency and increasing impacts. The HPC community has an opportunity to collectively leverage its expertise and the computing continuum to make a difference. One recent example is Hurricane Ottis. When hurricane Ottis recently made landfall in Acapulco, Mexico, as the first Category 5 storm ever to hit the Pacific Coast of North or South America, no formal hurricane warning had been issued. In fact, 16 hours before landfall, the National Hurricane Center still forecast only a Category 1 hurricane. 
%Within hours, it grew into a record-breaking, city-splintering Category 5 monster with winds of 165 miles per hour. 
Within hours, it grew into a record-breaking Category 5 hurricane with winds of 165 miles per hour. Ottis is probably the most expensive hurricane in Mexican history.  Experts analyzing the event and its progression have have noted that real-time sensor data along with the ability to integrate other parameters (e.g., water surface temperature and salinity) into the model could have been more effective in predicting this hurricane and its strength. Urgent computing approaches that leverage the computing continuum can have a tremendous impact in such events, in how we predict, detect, and manage our response, as well as how can we learn and adapt and evolve to do better.  

One initiative that is focused on establishing policies, structures, and mechanism for leveraging the computing continuum for support urgent applications is the US Whitehouse Office of Science and Technology Policy (OSTP) led \emph{National Strategic Computing Reserve (NSCR)}\footnote{The report, ``National Strategic Computing Reserve: A Blueprint,'' available at \url{https://www.nitrd.gov/national-strategic-computing-reserve-blueprint/}, outlines a Federal proposal for setting up a National Strategic Computing Reserve (NSCR) that can be called on in times of national crisis to rapidly activate a multi-sector advanced computing reserve infrastructure that can speed solutions, and it defines a blueprint for operational and coordination structures that will support an NSCR implementation.}\cite{friedlander2021u}. NSCR envisions an advanced computing cyberinfrastructure as a strategic national asset that can be mobilized during an emergency response. Its goal is to ensure the availability of a ready \emph{reserve} of resources (computing, data, software, services) and expertise that can be leveraged nimbly in times of urgent need, and to establish policies, processes, and agreements to enable effective resource mobilization and coordinate across agencies, stakeholder communities, and other national reserves. The NSCR vision is analogous to the roles of the US Civil Reserve Air Fleet and the United States Merchant Marine (among others) that can be called upon to assist the military in a crisis. Although this initiative is in its early stages, it can potentially transform how we  effectively leverage the computing continuum to detect and respond to national and global emergencies. 

\section{Acknowledgements}
The author acknowledges the contributions of Daniel Balouek and Ivan Rodero to the vision presented in the paper. The author is also grateful for the discussions and thoughtful feedback from a wide range of colleagues that have informed the ideas presented in this paper. 

\bibliographystyle{unsrtnat}
\bibliography{references.bib}

\end{document}